\documentclass[preprint,journal]{vgtc}       

\ifpdf
  \pdfoutput=1\relax                   
  \usepackage{graphicx}                
  \DeclareGraphicsExtensions{.pdf,.png,.jpg,.jpeg} 
\else
  \usepackage{graphicx}                
  \DeclareGraphicsExtensions{.eps}     
\fi%

\graphicspath{{figures/}{pictures/}{images/}{./}} 

\usepackage{microtype}                 
\PassOptionsToPackage{warn}{textcomp}  
\usepackage{textcomp}                  
\usepackage{mathptmx}                  
\usepackage{times}                     
\usepackage{cite}                      
\usepackage{tabu}                      
\usepackage{booktabs}                  
\usepackage{verbatim}
\usepackage{amsmath}
\usepackage{hyperref}
\hypersetup{
    colorlinks=true,
    linkcolor=blue,
    filecolor=blue,      
    urlcolor=blue,
    citecolor=cyan,
}

\preprinttext{VAST Challenge 2020: Honorable Mention for Streamlined Analysis Process.}

\onlineid{0}

\vgtccategory{Research}
\vgtcpapertype{please specify}

\title{%
  CA2: Cyber Attacks Analytics
}

\author{Luyu Cheng, Bairui Su, Yumeng Xue, Xiaoyu Liu, Yunhai Wang}
\authorfooter{
\item
 L. Cheng, B. Su, Y. Xue, X. Liu, and Y. Wang are with Shandong University, Qingdao. E-mail: \{chengluyu, pearminipro, xym6336, lxysduer, cloudseawang\}@gmail.com.
\item
 Y. Wang is the corresponding author.
}

\shortauthortitle{Biv \MakeLowercase{\textit{et al.}}: CA2: Cyber Attacks Analytics}

\abstract{The VAST Challenge 2020 Mini-Challenge 1 requires participants to identify the responsible white hat groups behind a fictional Internet outage. To address this task, we have created a visual analytics system named CA2: Cyber Attacks Analytics. This system is designed to efficiently compare and match subgraphs within an extensive graph containing anonymized profiles. Additionally, we showcase an iterative workflow that utilizes our system's capabilities to pinpoint the responsible group.} 


\renewcommand{\manuscriptnotetxt}{}

\vgtcinsertpkg

\begin{document}


\firstsection{Introduction}

\maketitle

The Center for Global Cyber Strategy has collected and constructed a graph containing profiles of the entire white hat community. The task is to identify the group that most closely resembles the provided template group among the five suspicious groups or all profiles. To accomplish this, we present a visual analytics system called CA2: Cyber Attacks Analytics, along with our solution for the given tasks.

\section{System Design}

The challenge involves a highly extensive graph, consisting of a template subgraph and five candidate subgraphs as input data. Each node within the graph represents one of the following entities: individuals, documents, demographics, and countries. Meanwhile, each edge indicates a specific communication channel, such as authorship, selling, purchasing, income and expenditure, as well as contact via phones or emails. As a result, our visual analytics system primarily concentrates on graph comparison and matching.

\subsection{Design Rationales}

The design of the system is based on following rationales.

\textbf{Comprehensive.} The system aims to cover as many aspects of the data as possible. This involves the exploitation of all potential data facets.

\textbf{Hierarchical.} The system organizes all components in a top-down hierarchy. This arrangement enables users to locate the specific components they wish to interact with by following the hierarchy.

\textbf{Coordinated Multiple Views.} Each component within the system is linked to its related components. This means that selections or modifications made by users in one chart are intended to be reflected in related charts.

\subsection{User Interface}

\begin{figure}[tb]
 \centering 
 \includegraphics[width=\columnwidth]{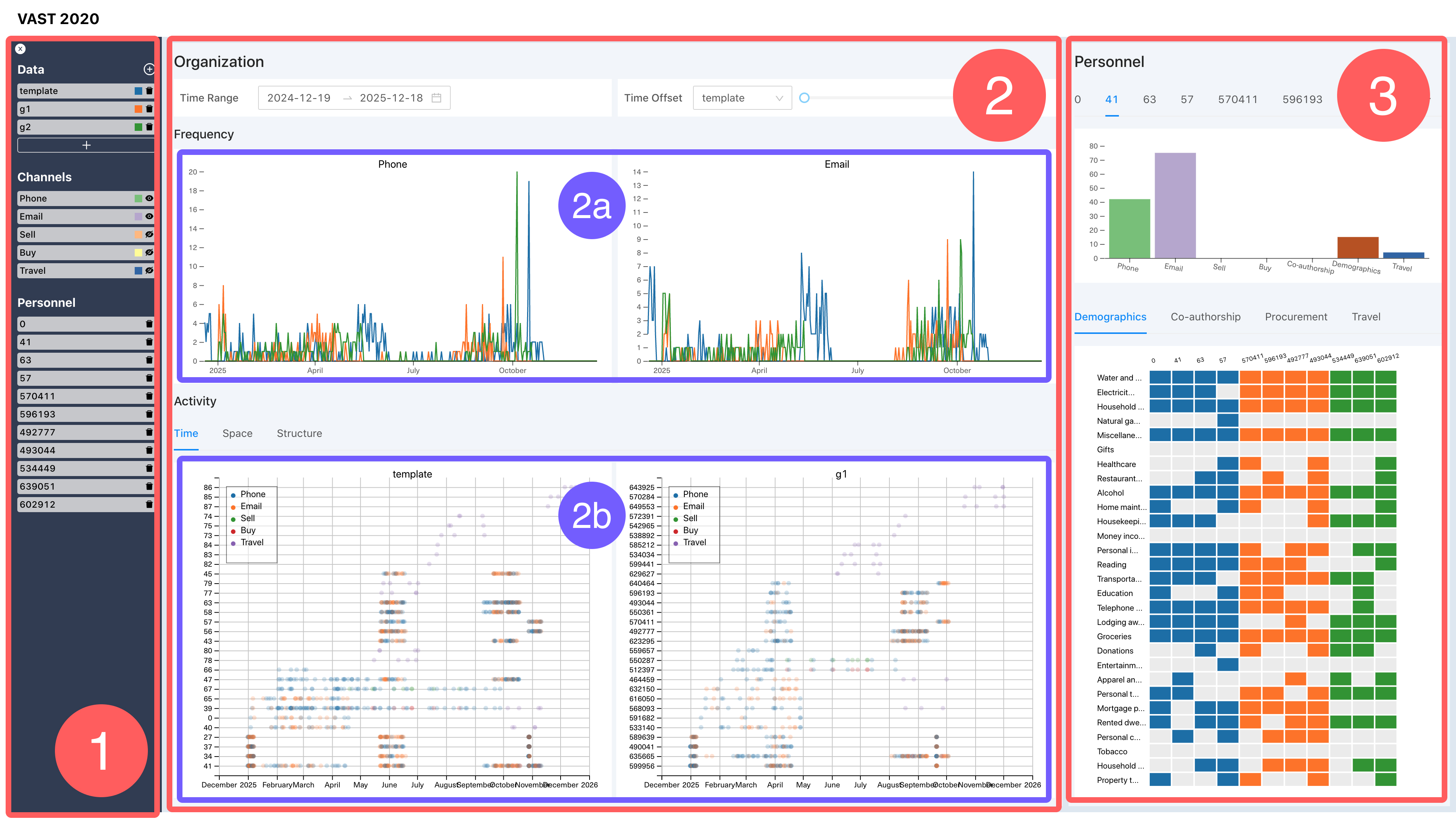}
 \caption{The system overview is composed of three main components: the control panel (1), the organization panel (2), and the personnel panel (3).}
 \label{fig:overview}
 \vspace{-0.35cm}
\end{figure}

Based on these principles, we designed and developed the user interface (Fig.~\ref{fig:overview}), which consists of three main sections. The leftmost section (Fig.~\ref{fig:overview}-1) is the control panel. The middle section (Fig.~\ref{fig:overview}-2) is the organization panel, housing charts related to personal relationships. The right section (Fig.~\ref{fig:overview}-3) is the personnel panel, which contains charts displaying personal data.

\subsubsection{The Control Panel}
The control panel serves two primary functions: 1) adding, removing, and replacing subgraphs; and 2) enabling or disabling channels. As a result, users are not confined to using only the provided template graph and candidate graphs; they can also import other subgraphs by uploading them. Additionally, users have the option to toggle the visibility of various channels across all components, allowing them to focus on the specific channels of interest.

\subsubsection{The Organization Panel}
The organization panel offers views that present information about relationships between individuals.
Users have the option to define a time range and time offset for each graph. This functionality enhances the user's ability to efficiently match each subgraph and identify potential subgraphs.
Regarding organization visualization, the system initially provides line charts (Fig.~\ref{fig:overview} (2a)) displaying the overall frequency of activities within each organization. This offers users an overview of each subgraph. In the lower section (Fig.~\ref{fig:overview} (2b)), three views are presented for visualizing activities, with each view focusing on a specific aspect.
\begin{figure}[tb]
 \centering
 \includegraphics[width=\columnwidth]{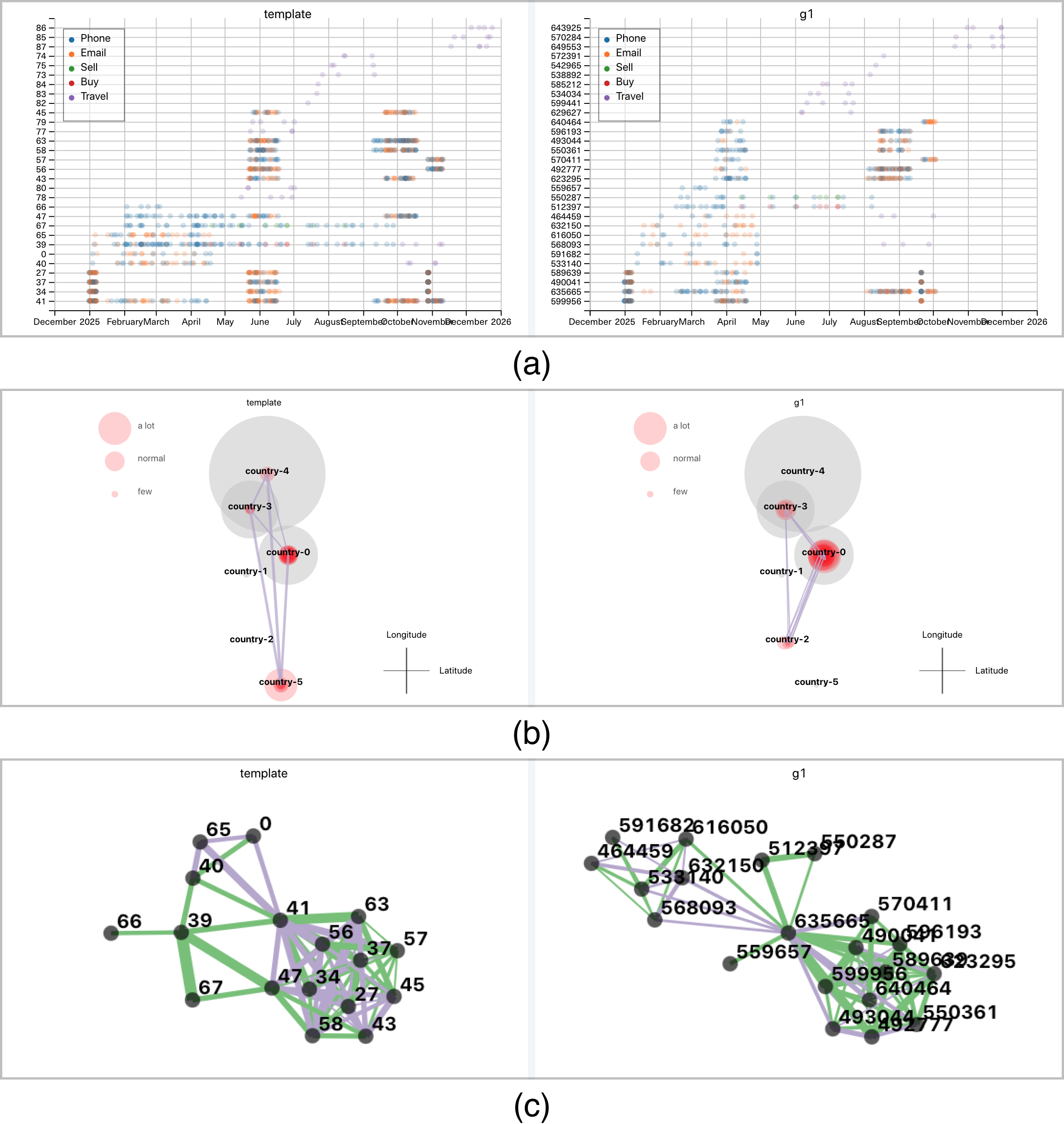}
 \caption{The organization panel includes three activity visualizations: (a) time view, (b) space view, and (c) structure view.}
 \label{fig:activity-visualizations}
 \vspace{-0.35cm}
\end{figure}

\textbf{Time View}. This view (Fig.~\ref{fig:activity-visualizations} (a)) offers a temporal scatterplot of the activities associated with each person within each organization. In this representation, each circle corresponds to an activity record. An interactive feature involves hovering the cursor over a circle, which triggers the highlighting of connections between the activity's owner and other individuals.

\textbf{Space View}. This view (Fig.~\ref{fig:activity-visualizations} (b)) offers a map for visualizing the spatial distribution of activities associated with each person within each organization. It's important to note that this map displays only the relative locations of individual countries, without indicating the precise latitude and longitude coordinates.

\textbf{Structure View}. This view (Fig.~\ref{fig:activity-visualizations} (c)) depicts the structure of each organization, with individuals connected based on the activities in which they participate. Users can click on nodes to select corresponding personnel. This feature proves highly beneficial in helping users discern the relationships between the template graph and specific subgraphs.

\subsubsection{The Personnel Panel}

\begin{figure}[tb]
 \centering 
 \includegraphics[width=\columnwidth]{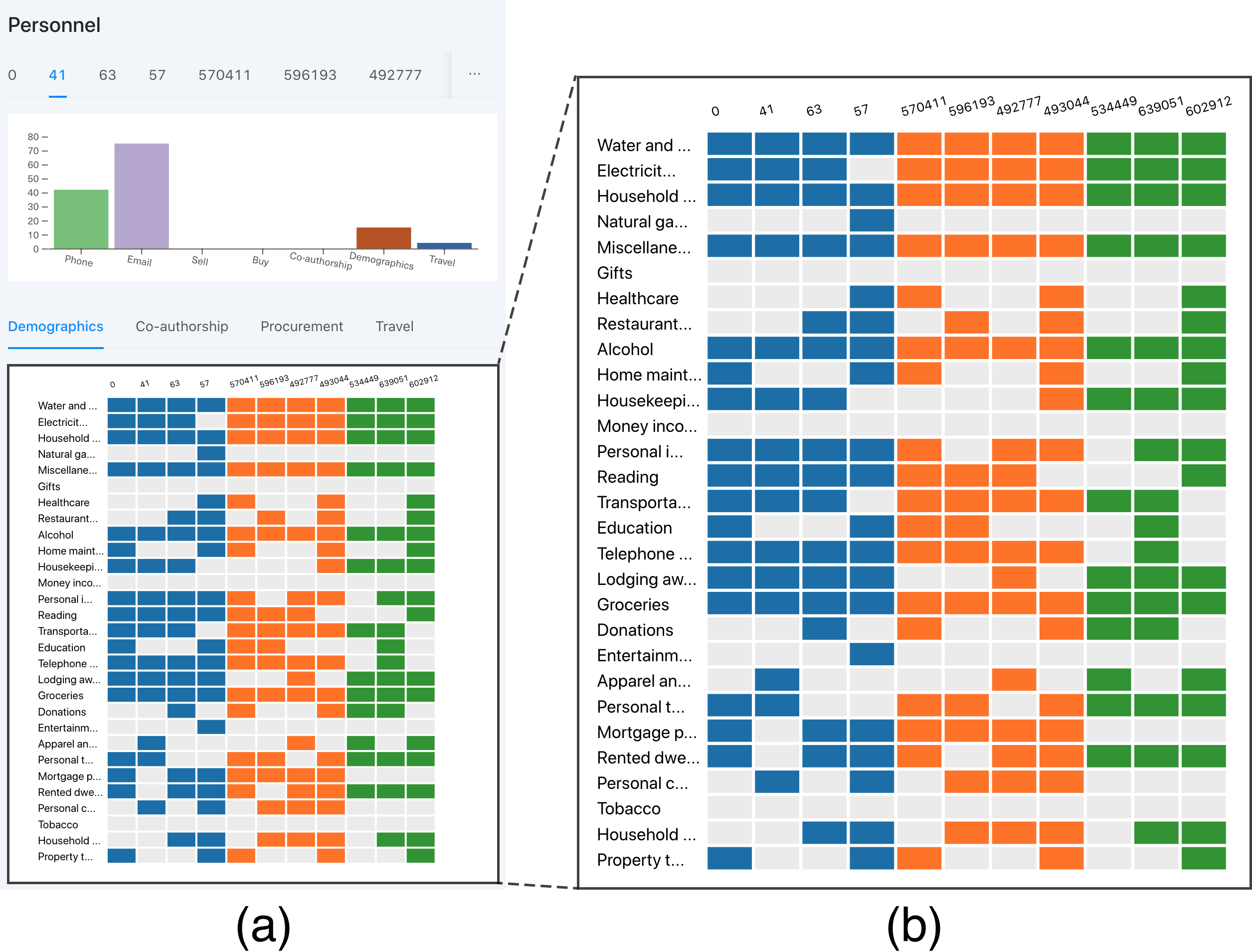}
 \caption{The personnel panel (a) of the system. Users can inspect selected persons' information in the top and comparing selected persons in terms of activities in the bottom (b).}
 \label{fig:personnel-panel}
 \vspace{-0.35cm}
\end{figure}

The personnel panel (Fig.~\ref{fig:personnel-panel}) is dedicated to providing information about the selected individuals. The top section features multiple tabs, each representing a selected person. Within each tab, a bar chart displays the number of activities associated with that person. In the lower section (Fig.~\ref{fig:personnel-panel} (b)), the tabs correspond to the channels selected in the control panel. In this section, a heatmap illustrates the activity frequency of all selected persons. By comparing the data across individuals, users can determine the similarities between selected persons from different subgraphs and facilitate matching.

\subsection{Implementation}

The system is constructed using web-page technology. The charts and interactive graphs are created with D3.js \cite{2011-d3} and the user interface is built with React \cite{react}. The system is available online at \texttt{\url{https://vast-2020.now.sh}} and the source code is accessible at \texttt{\url{https://github.com/pearmini/vast-2020}}.

\section{Large Graph Analysis}

Utilizing the system we presented, we successfully completed the task of comparing and matching the template graph with the provided candidate graphs. Moving forward, let's delve into how we can leverage the system to identify additional potential subgraphs within the extensive provided graph.

Considering the scale of each template subgraph, with over 100 million edges and more than 200 thousand vertices, it becomes necessary to filter out unrelated edges and vertices through matching. To achieve efficient retrieval, we've implemented a graph database for storing and searching the graph.

To locate matched nodes within the large graph, we conducted an analysis of the template graph using our visual analysis system. Based on our findings, we determined that only two individuals within the template graph are linked by procurement edges, and a solitary item node is associated with these two individuals. As a result, we employ the edge characteristics between these three nodes as criteria to identify matched nodes within the extensive graph. This leads us to propose an iterative workflow aimed at discovering new nodes from these matched nodes. By repeating these steps, users can uncover subgraphs similar to the template graph within the large graph.

\textbf{Step 1}. Formulate a set $S$ and put the matched nodes from the template into this set. Unmatched nodes are placed in a separate set $T$. The template graph $G_T = \langle  N_T, E_T\rangle  $, while the large graph is represented as $G_L = \langle  N_L, E_L\rangle  $.

\textbf{Step 2}. Establish a mapping $M$ to associate nodes in $N_T$ with their corresponding nodes in $N_L$.

\textbf{Step 3}. For each unmatched node $T_i \in T$ that has an adjacent node $A \in \operatorname{Adjacent}(T_i) \in S$, retrieve the node set $\beta = \operatorname{Adjacent}(M(A))$. For each $\beta_j \in \beta$, compare the similarity between $E\langle  A,T_i\rangle  $ and $E\langle  M(A),\beta_j\rangle  $, and identify the most similar node pair  $\langle  T_i, \beta_j\rangle  $.

\textbf{Step 4}. For each node peer $\langle  T_i, \beta_j\rangle  $ identified by \textbf{Step 3}, visualize the graph $\langle  S \cup T_i, E\langle  S,S\rangle   \cup E\langle  S,T_i\rangle  \rangle  $ and the graph $\langle  M(S) \cup \beta_i, E\langle  M(S),M(S)\rangle   \cup E\langle  M(S),\beta_j\rangle  \rangle  $. Then determined by the user whether it matches or not. If matched, update $M(T_i) = \beta_j$, move $T_i$ to set $S$, and remove it from set $T$.

\textbf{Step 5}. If set $T$ is empty, exit; otherwise, repeat \textbf{Step 3}.

\begin{figure}[tb]
 \centering 
 \includegraphics[width=\columnwidth]{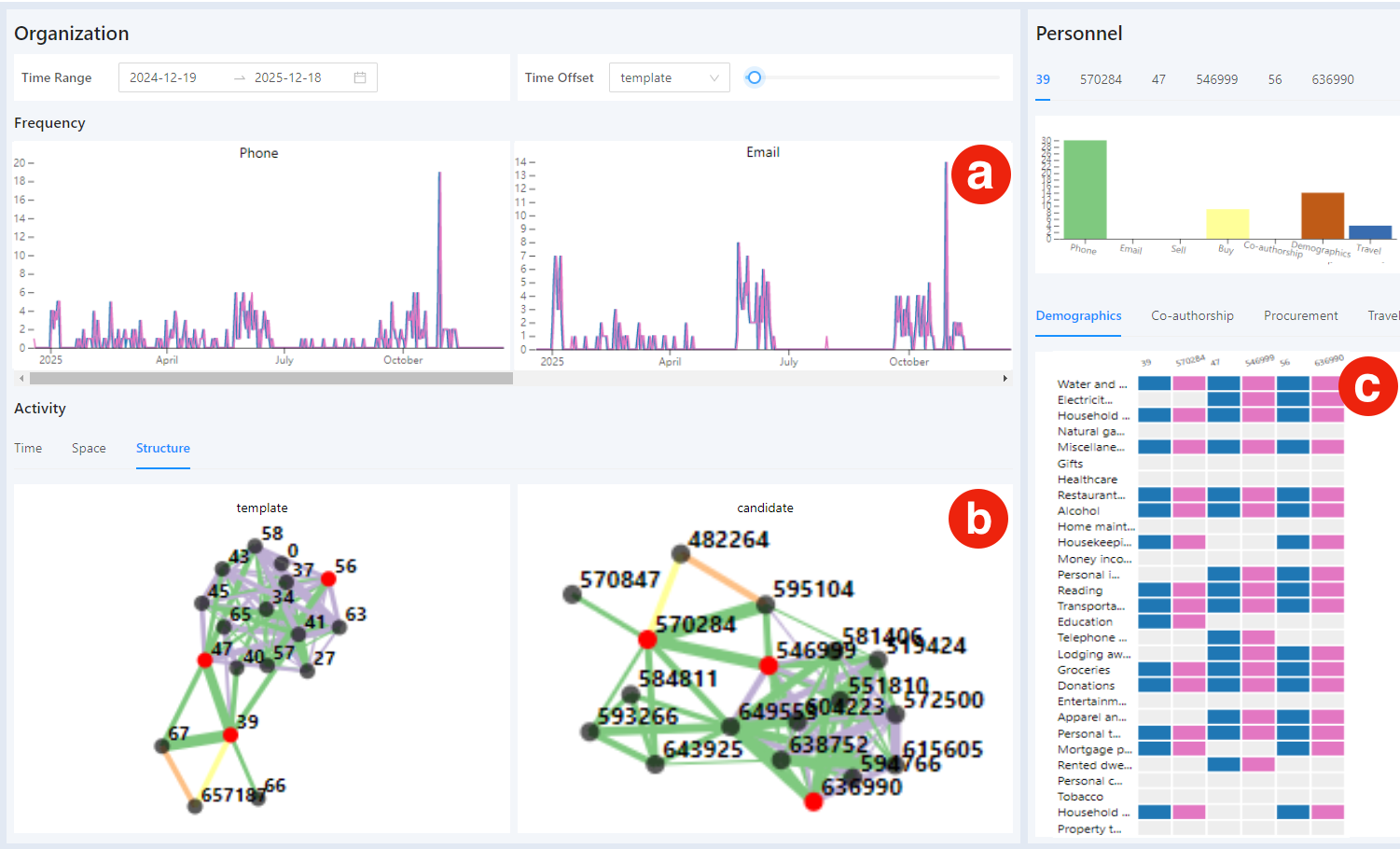}
 \caption{The result of large graph matching. The candidate graph and the template graph shows a high similarity in the frequency chart (a), the graph structure (b), and the heat map (c).}
 \label{fig:matching_result}
 \vspace{-0.35cm}
\end{figure}

As illustrated in Fig.~\ref{fig:matching_result}, the resulting subgraph closely resembles the template graph. We have included it in the online demo, accessible by adding a new subgraph labeled ``candidate'' in the control panel.


\section{Conclusion}

In this report, we introduce \textbf{CA2: Cyber Attacks Analytics}, a visual analytics system designed for comprehensive comparison and matching of the template subgraph with candidate graphs across various dimensions. This tool effectively facilitates the swift identification of the responsible white hat group, making it a valuable asset for addressing the requirements of VAST Challenge 2020 Mini-Challenge 1.


\bibliographystyle{abbrv-doi}

\bibliography{template}
\end{document}